\documentclass[useAMS,usenatbib]{mn2e}

\usepackage{amsfonts}
\usepackage{amsmath}
\usepackage{graphicx}

\title[The environmental dependence of clustering in hierarchical models]
{The environmental dependence of clustering in hierarchical models}

\author[U. Abbas and R. K. Sheth]
{Ummi Abbas$^{1}$\thanks{E-mail:  ummi@phyast.pitt.edu (UA); 
                                  shethrk@physics.upenn.edu (RKS)}
and Ravi K. Sheth$^{2}$\footnotemark[1]\\
$^{1}$Department of Physics \& Astronomy, University of Pittsburgh, 
      Pittsburgh, PA 15260, USA\\
$^{2}$Department of Physics \& Astronomy, University of Pennsylvania, 
      PA 19104, USA}

\newcommand{\bm}[1]{{\mbox{\boldmath $#1$}}}

\begin{document}
\pagerange{\pageref{firstpage}--\pageref{lastpage}}

\maketitle

\label{firstpage}

\begin{abstract}
In hierarchical models, density fluctuations on different scales are 
correlated.  This induces correlations between dark halo masses, their 
formation histories, and their larger-scale environments.  In turn, 
this produces a correlation between galaxy properties and environment.  
This correlation is entirely {\em statistical} in nature.  
We show how the observed clustering of galaxies can be used to quantify 
the importance of this statistical correlation relative to other 
physical effects which may also give rise to correlations between 
the properties of galaxies and their surroundings.  We also develop 
a halo model description of this environmental dependence of  
clustering.  
\end{abstract}



\section{\protect\bigskip Introduction}
A number of physical mechanisms are expected to play a role in 
determining the properties of a galaxy:  e.g., ram pressure stripping, 
harrassment, and strangulation 
~\citep{gunn72,farouki80,moore96,balogh00}.  
Many of these operate in dense environments.  
So the existence of a morphology--density 
relation---the fraction of galaxies which have elliptical rather than 
spiral morphologies is higher in denser regions \citep{dressler80}---is 
not unexpected.  For similar reasons, recent measurements of lower 
star-formation rates in denser regions, which appears to persist even 
at fixed morphology \citep{balogh02,gomez03} 
and stellar mass \citep{kauffmann04}, are not unexpected.  
However, determining which, if any, of the physical mechanisms 
mentioned above is the dominant one is more difficult.  

Hierarchical galaxy formation models have been rather successful at 
reproducing the morphology--density relation \citep[e.g.][]{benson01}.  
In these models, a correlation between galaxy-type and environment 
arises even if none of the physical mechanisms mentioned above are 
present.  The correlation is a consequence of the following assumptions.  
Gravity has transformed small fluctuations in the early Universe into 
the structures we see today.  This transformation was hierarchical, 
in the sense that small virialized objects formed first, and then merged 
with one another to form more massive virialized objects at a later time.  
The virialized objects present at any given time, called dark matter 
haloes, are approximately 200 times denser than the background universe 
at the time \citep{gunn72}.  
Galaxies form from gas which cools within virialized dark matter haloes 
\citep{white78}.  
The properties of a galaxy are determined entirely by the mass and 
formation history of the dark matter halo within which it formed 
\citep[e.g.][]{white91,kauffmann97,somerville99,cole00}.  
Halo masses and formation histories are directly related to the 
structure of the initial density fluctuation field from which 
they formed \citep*{press74,lacey93,shethmot01}. 
In hierarchical models, there is a 
correlation between fluctuations on different scales, and this 
induces correlations between halo mass and/or formation and the 
larger scale environment of a halo \citep{mowhite96,lemson99,shethtor02}. 
This, in turn, induces a correlation between galaxy-type and 
environment.  This correlation is entirely {\em statistical} in 
nature.  So it is interesting to ask if this statistical correlation 
is sufficient to explain most of the observed correlation between 
galaxy-type and environment.  

The main goal of the present work is to show how analysis of the 
clustering of galaxies can be used to quantify the strength of this 
statistical correlation between galaxy properties and environment.  
The idea is to show the strength of the statistical effect alone:  
any discrepancy with observations can then be ascribed to the other 
physical processes \citep{shethabbas04}.  
Although our argument is general, we focus in particular on the 
two-point correlation function, $\xi(r)$, and show how it is expected 
to depend on environment if the only environmental effects on galaxy 
properties are those which arise from the statistics of the initial 
fluctuation field.  The present analysis is intended to complement 
traditional measures of the environmental dependence of clustering 
which tend to focus on correlations between the distribution of an 
observable (e.g., the luminosity function, or the star formation rate, 
etc.) and the environment \citep[see e.g.][for some recent analyses]{yang03,
mo04,kauffmann04,balogh04,blanton04}.  

This paper is arranged as follows. 
Section~\ref{bigpicture} uses numerical simulations to illustrate how 
clustering is expected to depend on environment if the entire 
environmental depedence arises from the correlation between haloes and 
their environments.  
Section~\ref{sims} shows the effect for dark matter.  
A simple toy model which captures most of the relevant features of 
this effect is presented in Section~\ref{toymodel}.  
Section~\ref{mockgals} uses mock galaxy samples to show that the 
effect should be easily measured in surveys such as the SDSS.  
The final section summarizes our results and discusses some implications. 
An Appendix shows how to describe the effect using the language of 
the halo model \citep[see][for a review]{cooray02}
of large scale structure.  
 
\section{Environmental dependence of $\xi$}\label{bigpicture}
Throughout, we show results for a flat $\Lambda$CDM model for which 
$(\Omega_{0},h,\sigma_{8}) = (0.3,0.7,0.9)$ at $z=0$.  
Here $\Omega _{0}$ is the density in units of critical density today, 
the Hubble constant today is $H_{0}= 100 h $ km s$^{-1}$ Mpc$^{-1}$, 
and $\sigma_{8}$ describes the rms fluctuations of the initial 
field, evolved to the present time using linear theory, 
when smoothed with a tophat filter of radius $8h^{-1}$~Mpc.  
The GIF and VLS numerical simulations we use to illustrate our arguments 
were made available to the public by the Virgo consortium.  
Both were run with the same $\Lambda$CDM cosmology, but with 
slightly different parameterizations of the initial fluctuation 
spectrum.  The GIF simulation had $256^{3}$ particles in a cubic box 
with sides $L = 141h^{-1}$Mpc.  
The VLS simulation had $512^3$ particles in a cubic box with 
sides $L = 479h^{-1}$Mpc.  

\subsection{Dark matter simulations}\label{sims}
\begin{figure}
 \centering
 \includegraphics[width=\hsize]{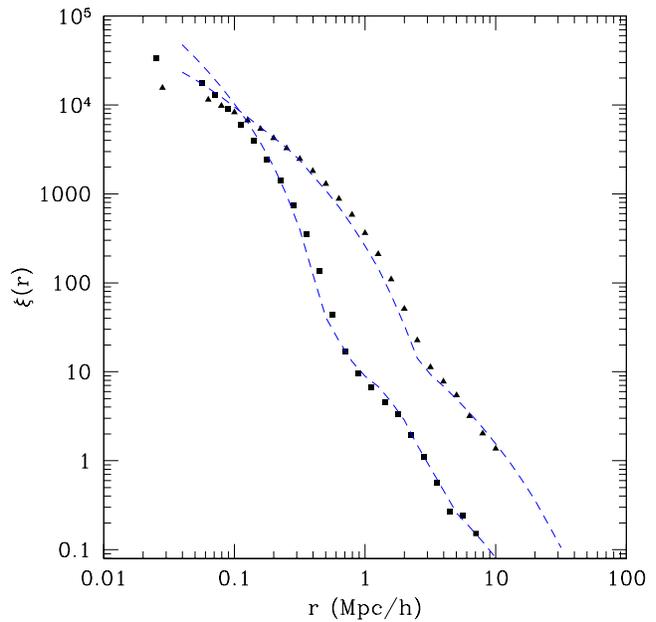}
 \caption{Environmental dependence of the dark matter 
 correlation function in the $\Lambda$CDM GIF simulation. 
 Triangles show $\xi(r)$ of particles in haloes which were defined 
 as having the densest environments (defined by counting the mass 
 within a sphere of radius of $5h^{-1}$Mpc centred on each halo), 
 and squares are from particles in underdense regions. Smooth curves 
 show the analytic model for this environmental effect that is developed 
 in Appendix~\ref{halomodel}.}
 \label{gifdm}
\end{figure}

To illustrate how clustering depends on environment, we used the halo 
and particle distributions in the $\Lambda$CDM GIF simulations 
\citep{kauffmann99}.  
The simulations contain approximately 90,000 haloes which each have 
at least ten particles, where the virialized halos were found using 
a friends-of-friends group finding program.  
We defined the environment of a halo by using the mass $M_R$ within 
a sphere of radius $R$ centred on the halo.  
We set $R=5h^{-1}$ or $8h^{-1}$Mpc, but any value which is 
substantially larger than the typical virial radius (a few hundred kpc), 
but smaller than the scale on which the Universe is essentially 
homogeneous, will do.  We then ranked all the haloes in decreasing 
order of $M_R$.  The particles belonging to the top one-third of the 
haloes were labeled as belonging to densest environments, and the 
particles in the bottom one-third of the halo sample were labeled 
as belonging to the least dense environments.  
Finally, we computed the correlation function of particles belonging 
to the haloes in the densest and least dense regions.  
The results are shown in Figure~\ref{gifdm}.  

There are obvious differences between the two correlation functions.  
The correlation function for the particles in dense regions extends 
to larger scales; if fit to a power-law, it would have a shallower 
slope.  The next section describes a simple model for these differences.  
The smooth curves in the figure show the result of a more complete 
analytic model that is developed in Appendix~\ref{halomodel}.  

\subsection{A toy model}\label{toymodel}
Let $dn(m,\delta_c)/dm$ denote the number density of dark matter haloes with 
mass $m$ at a time when the linear theory overdensity required for 
spherical collapse is $\delta_c$, and let $N(m,\delta_c|M,V)$ denote 
the average number of $m$ haloes in regions of volume $V$ which contain 
mass $M$.  
Define $M/V\equiv \bar\rho (1+\delta)$, where $\bar\rho$ denotes the 
average density of the background.  Dense regions have $\delta>0$.  
\cite{mowhite96} showed that a generic feature of hierarchical 
models is that 
$N(m,\delta_c|M,V)\ne (1 + \delta)\, V\, dn/dm$:  i.e., dark halo 
abundances in dense and underdense regions do not differ by a simple 
factor which accounts for the difference in density.  Rather, 
\begin{equation}
 N(m,\delta_c|M,V) \approx 
  \Bigl[1 + b(m)\delta\Bigr] V{dn(m,\delta_c)\over dm},
 \label{nmdelta}
\end{equation}
where 
\begin{equation}
 b(m,\delta_c)\approx 1-{d\ln dn(m,\delta_c)/dm\over d\delta_c}
 \label{bias-taylor}
\end{equation}
is a function which typically increases monotonically with $m$ 
\citep[e.g.][]{shethtor99}.  As a result, one expects the ratio of 
the number of massive to low mass haloes to be larger in dense regions 
than in less dense regions:  the mass function in overdense regions 
should be `top-heavy'.  Measurements in simulations indicate that this 
is indeed the case \citep[e.g.][]{lemson99,shethtor02}:  
the average halo mass increases with $\delta$.  

In the GIF simulations, the average mass of the $\sim 30,000$ haloes 
which reside in the densest regions is approximately 
$m=2.5\times 10^{12}h^{-1}M_\odot$, 
whereas the average mass of the $\sim 30,000$ haloes which reside in 
the least dense regions is $m=5.2\times 10^{11}h^{-1}M_\odot$.  
(Hence, the two overdensities differed by a factor of approximately 5.)

The fact that dense regions have a top-heavy mass function has an 
important consequence for the environmental dependence of the 
correlation function which we now describe.  
Let $\xi(r|\delta)$ denote the shape of the correlation function in 
regions with overdensity $\delta$.  Although such regions contain 
haloes with a range of masses, suppose we require that all haloes 
have the same mass $m_\delta$, chosen to match, say, the mean halo 
mass in the regions.  If these haloes are not clustered, then 
$\xi(r|\delta)$ is simply a consequence of the shape of the density 
profiles $\rho(r|\delta)$ around haloes \citep{peebles74}: 
\begin{equation}
 \xi(r|\delta) = n_{clus}(m_\delta) 
            \int d{\bm s}\,{\rho({\bm s}|m_\delta,\delta)\over\bar\rho_\delta} 
               {\rho({\bm s}+{\bm r}|m_\delta,\delta)\over\bar\rho_\delta} ,
\end{equation}
where $n_{clus}(m_\delta)$ is the average number density of haloes 
surrounded by regions with overdensity $\delta$, 
$\bar\rho_\delta \equiv m_\delta n_{clus}$, and 
$m_\delta\equiv \int d{\bm s}\,\rho({\bm s}|m_\delta,\delta)$.  
In fact, the haloes are clustered, but \cite{shethjain97} show why 
ignoring halo clustering should be a good approximation on small scales.  
Hence, to estimate the small scale correlations as a function of 
environment, we require an estimate of the shapes of halo density 
profiles.  

When fit to spherical models, the density profiles of haloes in 
simulations are well fit by the functional form
 $\rho(r|m)/\bar\rho=\Delta_{c} (r_{vir}/cr)/(1 + cr/r_{vir})^2$ 
where $r_{vir} \equiv (3m/4\pi\Delta_{vir}\bar\rho)^{1/3}$ 
and $c\equiv 9\,(10^{13}h^{-1}M_\odot/m)^{0.1}$ \citep{navarro97}.  
If we assume that the density profiles of $m$ haloes are the same in 
all environments (we will modify this assumption shortly), then 
$\xi(r|\delta)$ is given by inserting this expression for the density 
profile into the convolution integral above.  The result is 
\begin{equation}
 \xi(r|\delta) = {\Delta_{vir}\over \bar\rho_\delta/\bar\rho}\ 
                 \lambda(r|m_\delta,\delta),
 \qquad {\rm provided}\quad r\le 2r_{vir},
 \end{equation}
where $\lambda$ is a messy function of $c$ and $cr/r_{vir}$ 
\citep[given in][]{sheth01}.  
Since $\Delta_{vir}$ is independent of $m$, in such a model, the 
environmental dependence of $\xi(r|\delta)$ comes entirely from the 
fact that dense regions host the more massive haloes, and halo density 
profiles depend on halo mass.  The factor of $\bar\rho_\delta/\bar\rho$ 
in the denominator derives from the fact that haloes which have a fixed 
overdensity relative to the global background density $\bar\rho$ have 
a different overdensity relative to the local background 
$\bar\rho_\delta$.

\begin{figure}
 \centering
 \includegraphics[width=\hsize]{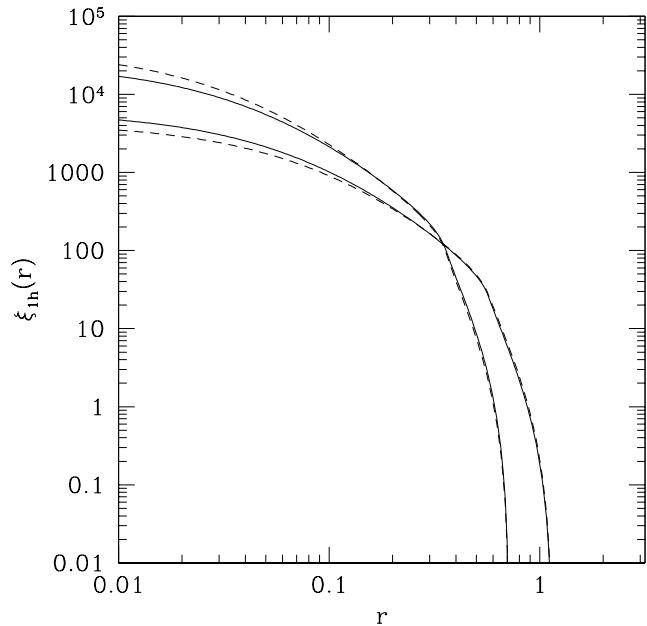}
 \caption{The correlation function of particles in haloes which are 
 surrounded by dense regions (curves which extend to larger $r$), 
 and by less dense regions (curves which fall to zero at smaller $r$), 
 in our toy model.  
 The number of haloes in the two environments was assumed to be the 
 same, but the halo mass in dense regions was assumed to be larger by 
 a factor of 5.  Solid curves have $c = 10$ in both cases, whereas 
 dashed curves include a realistic prescription for the weak mass 
 dependence of $c$.}
 \label{simple}
\end{figure}

Figure~\ref{simple} shows the result of this simple analytic estimate 
of $\xi(r|\delta)$ for the two sets of GIF simulation particles:  
those which reside in the 30,000 haloes with the largest surrounding 
overdensities (as described previously), and those which reside in the 
30,000 haloes with the smallest surrounding overdensities 
The curves are qualitatively similar to the measurements shown in 
Figure~\ref{gifdm}, at least out to scales on which the measurements 
show an inflection:  $\xi$ falls to zero on smaller scales in the 
less dense regions.  
The inflections at $\sim 0.8h^{-1}$Mpc and $\sim 3h^{-1}$Mpc in 
the simulations (Figure~\ref{gifdm}) denote the scales which are 
approximately twice the virial radii of the typical haloes in the two 
regions.  Beyond this scale, halo-halo correlations become important; 
we build a model for this in Appendix~\ref{halomodel}.

The curves which extend to larger scales are those for particles in 
the denser regions.  This is easily understood, because dense regions 
are those for which $m_\delta$ is larger, and hence $r_{vir}$ is larger.  
The reason why $\xi(r|\delta)$ on small scales is larger for the less 
dense regions is more subtle.  

The solid curves show results in which we have set $c=10$ and ignored 
the mass dependence of $c$, and dashed curves include the mass 
dependence but assume that there is no additional dependence on 
environment.  Clearly, the mass dependence of $c$ is not a dominant 
effect even on scales smaller than $r_{vir}/c$.  Thus, the 
difference in amplitudes on small scales derives from the factor of 
$\bar\rho_\delta/\bar\rho$ in the expression above, and not from the 
mass dependence of $c$.  In the present example, the number of haloes in 
the two (dense and underdense) regions is the same, but $m_\delta$ in 
dense regions is larger, so $\bar\rho_\delta/\bar\rho$ is larger for 
the dense regions.  Since the shape of $\xi$ is approximately constant 
on small scales, it has a lower amplitude in denser regions.  

The relative unimportance of $c$ has the following interesting 
consequence.  Suppose that halo density profiles do depend on 
environment (numerical simulations are only just reaching the 
resolution required to address this question).  
A simple way of parameterizing this dependence is to allow $c$ to depend 
both on $m$ and $\delta$.  If $c$ depends only weakly on $\delta$, 
then the effect on $\xi$ will only be noticeable on scales smaller than 
$r_{vir}/\bar c$, where $\bar c$ denotes the mean $c$ (averaged over 
environments).  

On the smaller scales, where halo-halo correlations are not important, 
the differences between the two curves in Figure~\ref{gifdm} are 
qualitatively like those of the simple toy model described above, 
indicating that our use of a mean density-dependent halo mass $m_\delta$, 
rather than a distribution of masses, does capture the essential 
features of the density dependence of $\xi(r|\delta)$.  
On larger scales, where halo correlations are important, 
Figure~\ref{gifdm} shows that $\xi(r|\delta)$ is stronger in dense 
regions.  This is not unexpected in the context of the linear 
peaks-bias model of \citep{kaiser84}, if, on average, the densest regions 
at the present time formed from the densest regions in the initial 
fluctuation field.  
This is because, in the initial Gaussian random field, the densest 
regions were more strongly clustered than regions of average density.  
Therefore, our simple model, in which the environmental dependence of 
$\xi$ is entirely due to the environmental dependence of the halo mass 
function, suggests the following generic features for $\xi(r|\delta)$:  
on scales larger than the virial radius of a typical halo, the 
amplitude of $\xi$ should increase as $\delta$ increases; 
on smaller scales, the amplitude of $\xi$ in underdense regions 
should be larger; 
hence, the slope of $\xi$ in dense regions should be shallower 
than in less dense regions.

\subsection{Mock galaxy samples}\label{mockgals}
To illustrate that the features described above really are generic, 
and to make a closer connection to observations, we assigned mock 
`galaxies' to haloes in the simulations as follows.  
Sufficiently low mass haloes contain no galaxies.  
Haloes more massive than some $m_L$ contain at least one galaxy.  
The first galaxy in a halo is called the `central' galaxy.  
The number of other `satellite' galaxies is drawn from a Poisson 
distribution with mean $N_s(m)$ where 
\begin{equation}
 N_s(m) = \left(\frac{m}{m_1} \right)^\alpha  \text{ if }m \geq m_L.  
 \label{Ngsdss}
\end{equation}
This procedure is motivated by \cite{kravtsov04}.  
We distribute the satellite galaxies in a halo around the halo centre 
so that the radial profile follows that of the dark matter (i.e., the 
galaxies are assumed to follow an NFW profile).  
We set $m_L=10^{11.27}h^{-1}M_\odot$, $m_1=23m_L$, and 
$\alpha=0.92$; \cite{zehavi04} show that this choice is 
appropriate for SDSS galaxies more luminous than $M_r<-18$.  
We then compute $N_5$, the number of galaxies in a $5h^{-1}$Mpc 
sphere around each galaxy, and rank the galaxies in order of 
decreasing $N_5$.  The top one-third are labeled as being galaxies 
in dense regions, and the bottom third as being in underdense regions.  

\begin{figure}
 \centering
 \includegraphics[width=\hsize]{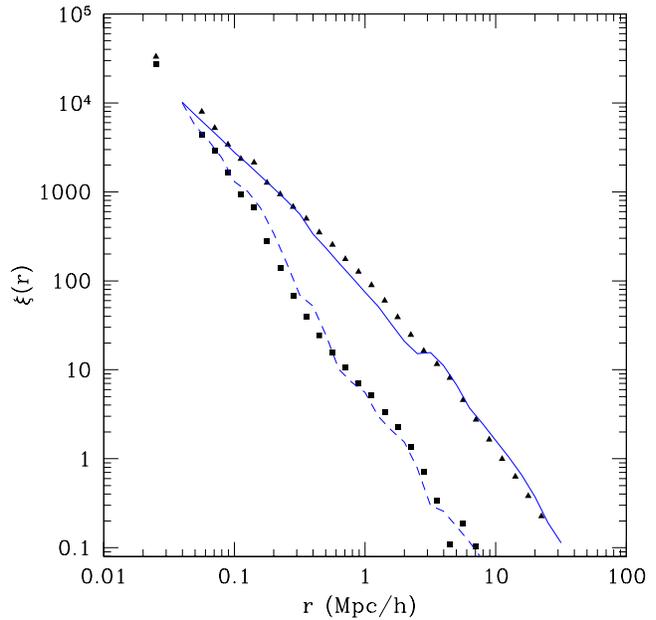}
 \caption{Similar to Figure~\ref{gifdm}, but now for model galaxies 
 distributed according to the model described in the text around 
 equation~\ref{Ngsdss}, and density defined by counting galaxies 
 in spheres centred on the galaxies themselves, rather than the mass 
 in spheres centred on haloes.  Curves show the analytic model 
 developed in Appendix~\ref{halomodel}.}
\label{gifsdss}
\end{figure}

Figure~\ref{gifsdss} presents the correlation functions of the mock 
galaxies classified as being in dense (top) and underdense (bottom) 
regions.  Once again we see the generic tendency for $\xi$ computed 
using objects in denser regions to be shallower than when objects in 
sparser environments are used.  

The important point, which we note explicitly here, is the following:  
By assuming that equation~(\ref{Ngsdss}) is the same function of $m$ 
for all environments, and by assuming that the radial profile of the 
galaxies depends only on halo mass and not on environment, we have 
constructed a galaxy catalog in which all environmental effects are 
{\em entirely} a consequence of the correlation between halo mass and 
environment.  Therefore, the locii traced out by the two sets of 
symbols shown in Figure~\ref{gifsdss} represent the predicted 
environmental dependence of $\xi$ if there are no environmental 
effects other than the statistical one determined by the initial 
fluctuation field.  

Over the $1-10h^{-1}$Mpc range which the SDSS data currently probe 
most reliably, $\xi$ for the dense and underdense samples differs by 
an order of magnitude.  This difference is easily measurable in data 
sets which are currently available.  
Comparison of this predicted difference with measurements in the SDSS 
will provide a sharp test of the assumption that environmental effects 
are dominated by the statistical correlation between the halo mass 
function and environment, rather than by other `gastro'physics.

\section{Discussion and Conclusions}
In hierarchical models, the clustering of dark matter particles should 
be a strong function of environment.  This is a consequence of the 
fact that massive haloes populate the densest regions.  
If the properties of galaxies are determined entirely by the masses 
and formation histories of the haloes in which they sit, then the 
clustering of galaxies should also depend strongly on environment.  

We discussed a method for testing this assumption.  The test is 
particularly interesting given recent work which suggests that, 
at fixed mass, haloes in dense regions form at higher redshifts 
\citep{shethtor04}, and that this effect is a decreasing 
function of halo mass \citep{gao05}.  In particular, 
Section~\ref{mockgals} described how to generate a mock galaxy catalog 
in which all correlations with environment are a consequence of the 
correlation between halo abundances and environment.  Comparison of 
the environmental dependence of clustering in the mock catalog and 
in real data allows one to test the assumption that environmental 
effects are dominated by the statistical correlation between halo 
mass and environment.  

The Appendix shows how to incorporate the assumption that environmental 
effects are determined entirely by the statistical correlation between 
halo mass and environment into the halo model description of large 
scale structure.  While this model aids in understanding the effect, 
to perform the test with data, it is unnecessary---measurements of 
clustering in the mock catalogs of the sort described in 
Section~\ref{mockgals} are sufficient.  
However, the methodology developed in this work provides a means for computing 
the clustering properties of galaxies affected by other environment 
dependent processes (assuming that some model for 
$\langle N_{gal}|m,M,V\rangle$ could be derived) 
that explicitly do not depend upon the surrounding halo mass. 
For instance, winds from galaxies in nearby clusters
are assumed to depend on the larger scale environment and could
affect the clustering statistics.

In particular, on scales larger than a Megaparsec or so, the 
two-point correlation function $\xi$ of galaxies surrounded by 
high density regions is predicted to be larger than for galaxies in 
less dense regions.  In addition, the slope of $\xi$ for galaxies in 
dense regions should be shallower (cf. Figure~\ref{gifsdss}).  
And if the distribution of galaxies around halo centers depends only 
weakly on environment (current analyses of galaxy clustering assume 
any such dependence is negligible), the effect on their clustering 
amplitude is small.  

Our test using mock catalogs is unrealistic in one respect:  
with real data, one only has redshift-space positions, so one cannot 
define the local density in real-space.  However, if the scale on 
which the redshift-space density is defined is larger than the typical 
size of redshift-space distortions ($\sim 5h^{-1}$Mpc), then the 
real and redshift-space densities will not differ substantially.  
Hence, we do not expect our separation into environments based on the 
real-space density within $8h^{-1}$Mpc spheres to differ substantially 
from that which we would have obtained had we used redshift-space 
positions instead.  A more detailed description of the redshift space 
method is the subject of work in progress.  (E.g., what does one gain 
by measuring both $\xi(s)$ and the projected correlation function as 
a function of redshift space environment?  The latter quantity 
depends on local density for similar reasons that $\xi(r)$ does---the 
halo mass function depends on environment---whereas $\xi(s)$ will 
have an additional effect coming from the fact that redshift space 
distortions depend on halo mass.)  
Comparison of these predictions with clustering in the 
Sloan Digital Sky Survey is underway.

\section*{Acknowledgements}

We thank the Virgo consortium for making their simulations available 
to the public, and the Pittsburgh Computational Astrostatistics group (PiCA)
for the NPT code which was used to measure the correlation functions 
in the simulations. We also thank an anonymous referee for suggesting minor
changes that helped to improve the paper.
This material is based upon work supported by the 
National Science Foundation under Grants No. 0307747 and 0520647.

\appendix

\section{An analytic model}\label{halomodel}
This Appendix provides an analytic model which incorporates the 
assumption that environmental effects are determined entirely by the 
statistical correlation between halo mass and environment.  
Note that to perform the test with data, this analytic model is 
unnecessary---measurements of clustering in the mock catalogs 
described previously are sufficient.  

The analysis which follows uses the framework of the halo model of 
large scale structure \citep[see][for a review]{cooray02}.  
It extends that of the toy model described in the main text in two ways:  
it allows for a range of halo masses, and it allows for correlations 
between haloes.  

\subsection{The halo model for dark matter}\label{standard}
In this model, all mass is bound up in dark matter haloes which have a 
range of masses.  Hence, the background density is 
\begin{equation}
 \bar{\rho} = \int dm \, {dn(m)\over dm}\, m.
 \label{rhobar}
\end{equation}
where $dn(m)/dm$ denotes the number density of haloes of mass $m$.  
The correlation function is the Fourier transform of the power spectrum 
$P(k)$:
\begin{equation}
 \xi(r) = \int {dk\over k}\, {k^3 P(k)\over 2\pi^2}\, {\sin kr\over kr} .
 \label{xir}
\end{equation}
In the halo model, $P(k)$ is written as the sum of two terms: 
one that arises from particles within the same halo and dominates 
on small scales (the 1-halo term), 
and the other from particles in different haloes which dominates 
on larger scales (the 2-halo term).  Namely, 
\begin{equation}
 P(k) = P_{1h}(k) + P_{2h}(k),
\end{equation}
where 
\begin{equation}
 \begin{split}
   P_{1h}(k) &= \int dm\,{dn(m)\over dm}\,\frac{m^2}{\bar{\rho}^2}\,
                     |u(k|m)|^2 , \\
   P_{2h}(k) &= \left[\int dm\,{dn(m)\over dm}\, \frac{m}{\bar{\rho}}\, 
                     b(m)\,u(k|m)\right]^2 P_{\rm Lin}(k).
 \end{split}	
 \label{Pk1h2h}
\end{equation}
Here $u(k|m)$ is the Fourier transform of the halo density profile, 
$b(m)$ is the bias factor which describes the strength of halo 
clustering, 
and $P_{Lin}(k)$ is the power spectrum of the mass in linear theory.
When explicit calculations are made, we assume that the density profiles 
of haloes have the form described by \cite{navarro97}, 
and that halo abundances and clustering are described by the 
parameterization of \cite{shethtor99}:  
\begin{eqnarray}
 {m\over\bar\rho}\,{dn(m)\over dm}\,{\rm d}m &=& 
  f(m)\,{\rm d}m = f(\nu)\,{\rm d}\nu \nonumber\\
   &=& {{\rm d}\nu^2\over \nu^2}\,
   \sqrt{a\nu^2\over 2\pi} \, \exp\left(-{a\nu^2\over 2}\right)
   A\,\left[1 + (a\nu^2)^{-p}\right] \nonumber\\
 b(m) &=& 1 + {a\nu^2 - 1\over\delta_{\rm sc}} 
            + {2p/\delta_{\rm sc}\over 1 + (a\nu^2)^p}
\nonumber\\
 \nu  &=& {\delta_{\rm sc}\over\sigma(m)}\quad {\rm and} \nonumber\\ 
 \sigma^2(m) &=& \int_0^\infty {{\rm d}k\over k}\,
                               {k^3P_{\rm Lin}(k)\over 2\pi^2}\ W^2(kR_0),
 \label{st99}
\end{eqnarray}
where $W(x)=(3/x^3)[\sin(x) - x\cos(x)]$ and $R_0=(3M/4\pi\bar\rho)^{1/3}$.  
That is to say, $\sigma(M)$ is the rms value of the initial fluctuation 
field when it is smoothed with a tophat filter of comoving size $R_0$, 
extrapolated using linear theory to the present time.  
Here $\delta_{\rm sc}$ is the critical density required for spherical 
collapse, extrapolated to the present time using linear theory (it is 
1.686 for an Einstein de-Sitter model), and $a\approx 0.71$, $p=0.3$ 
and $A = (1 + \Gamma(1/2-p)/\sqrt{\pi}/2^p)^{-1}\approx 0.322$.  
If $a=1$, $p=0$ and $A=1/2$, then ${\rm d}n/{\rm d}m$ is the same as 
the universal mass function first written down by \cite{press74}.  

\subsection{Including the environmental effect}\label{modelI}
The expressions above are the result of averaging over environments.  
Including the environmental dependence of the halo distribution 
explicitly is not entirely straightforward, because, as we 
describe below, it requires knowledge of the probability that a 
region of volume $V$ has overdensity $\delta$.  As we describe 
below, we use the excursion set model described in \cite{sheth98} to 
do this.  

In this model, spherical evolution is described by a `moving barrier':
\begin{equation}
  \delta_0(\Delta) = {\delta_{\rm sc}\over 1.686}\,
       \Biggl[1.686 -{1.35\over\Delta}-{1.124\over\Delta^{1/2}} 
       + {0.788\over\Delta^{0.587}}\Biggr]
\end{equation}
where $\Delta\equiv M/\bar\rho V$.  This barrier is said to be `moving' 
because, for general $V$, it is a function of $M$.  The excursion set 
model attributes special significance to the first crossing 
distribution $f(M,V)dM$ of this barrier by Brownian motion random walks: 
it is a measure of the mass fraction in regions of size $V$ which 
contain mass $M$.  
In this approach, a halo can be thought of as a patch which has 
collapsed to vanishingly small volume.  
In the limit of $V\to 0$, $\delta_0\to\delta_{\rm sc}$ is the same 
constant for all $M$.  In this limit the barrier is said to have 
constant `height', and the first crossing distribution 
$f(M,\delta_{\rm sc})dM$ represents the mass fraction in haloes of mass 
$M$.  Thus, $f(M,\delta_{\rm sc})$ equals $(M/\bar\rho)$ times 
$dn(M,\delta_{\rm sc})/dM$, the halo mass function.  

Notice that, for general $V$, $\delta_0(\Delta)\le\delta_{\rm sc}$ 
for all $M$.  
Hence, $f(m,\delta_{\rm sc}|M,V)$, the first crossing distribution 
of $\delta_{\rm sc}$ by walks which first crossed the moving barrier 
associated with non-zero $V$ at $M$, denotes the mass fraction of cells 
of size $V$ which contain mass $M$ which is in haloes which have mass 
$m$ for some $m\le M$.  
This fraction equals $(m/M)$ times $N(m,\delta_{\rm sc}|M,V)$, 
the environmental dependent mass function discussed in the main text 
(cf. equation~\ref{nmdelta}).  

In the main text (e.g. Section~\ref{sims}), we define the environment of 
a halo by considering the mass in a patch of volume $V$ surrounding it.  
To classify haloes by their environment, we must be able to estimate the 
number density of haloes of mass $m$ which are surrounded by regions of 
volume $V$ which contain mass $M$.  
If $dn(m)/dm$ denotes the number density of $m$ 
haloes, and $f(M,V|m)$ denotes the fraction of $m$ haloes which contain 
mass $M$ in the surrounding volume $V$, then the number density of 
such haloes is $f(M,V|m)\,dn(m)/dm$.  
In the excursion set model, this equals 
\begin{equation}
\begin{split}
 {dn(m)\over dm}\,f(M,V|m) dM &= {\bar\rho\over m}\,f(m)f(M,V|m)\,dM\\
	&= \frac{\bar{\rho}}{m}\,f(M,V)f(m|M,V)\,dM \\
	&= n(M,V)\,N(m|M,V)\,dM,
\end{split}
\end{equation}
where $f(M,V)dM$ is given by computing the first-crossing distribution 
of the moving barrier associated with spherical collapse described 
above.  \citep[In practice, we use the analytic approximation to the first 
crossing distribution of such moving barrier problems given by][]{shethtor02}.  

The mass density contributed by haloes that are embedded in regions of 
mass $M_{min}\le M\le M_{max}$ is;
\begin{eqnarray}
  \bar{\rho}_\delta &=& \int_{M_{min}}^{M_{max}}dM 
	         \int_0^{M_{max}}dm\, {dn(m)\over dm}\,f(M,V|m)\,m \nonumber\\
     &=& \int_{M_{min}}^{M_{max}} \!\!\!\!\!\!\! dM\,n(M,V) 
	                        \int_0^M dm\, N(m|M,V)\,m \nonumber\\
     &=& \int_{M_{min}}^{M_{max}}\!\!\!\!\!\!\! dM\, n(M,V)\,M .
\end{eqnarray}

In the standard model, the density profile of a halo depends on its mass, 
but not on the surrounding environment. In this case, the one-halo term is
\begin{equation}
 \begin{split}
   P_{1h}(k|\delta) = & \int_{M_{min}}^{M_{max}} dM\, n(M,V)\\ 
    \ & \times \int_0^M dm \,
   N(m|M,V)\left(\frac{m}{\bar{\rho_\delta}}\right)^2 |u(k|m)|^2.
 \end{split}
\end{equation}
This reduces to the standard 1-halo term in the limits 
 $M_{min} \rightarrow 0$ and $M_{max}\to\infty$.  

The two-halo term is more complex as it now has two types of 
contributions: pairs which are in the same patch (2-halo--1-patch),
and pairs in different patches (2-halo--2-patch).  
The 2-halo--1-patch term can only be important on intermediate scales 
(i.e., those which are larger than the diameter of a typical halo but 
smaller than the diameter of a patch).  It is more complex to model 
this term accurately, as we describe shortly.  The 2-halo--2-patch 
term, on the other hand, is simpler.  It should be well approximated by 
\begin{eqnarray}
 \begin{split}
 {P_{2h-2p}(k|\delta)\over P_{\rm Lin}(k|R_p)} &= 
       \Biggl[\int_{M_{min}}^{M_{max}}dM\, n(M,V)\,B(M,V) \\
    \ &\qquad\times \int_0^M \!\! dm\,
           N(m|M,V)\frac{m}{\bar{\rho_\delta}}\,u(k|m)\Biggr]^2 ,
 \end{split}
 \label{Pk2h2p}
\end{eqnarray}
where $P_{\rm Lin}(k|R_p)$ denotes the power spectrum associated 
with setting the linear theory correlation function to $-1$ on 
scales smaller than the diameter of a patch $2R_p$.  This 
truncation has little effect on small $kR_p\ll 1$, where 
$P_{\rm Lin}(k|R_p)\approx P_{\rm Lin}(k)$.  
The factor $B(M,V)$ describes bias associated the clustering of the 
patches, and depends on the abundance of such patches in exactly 
the same way that $b(m)$ is related to $dn(m)/dm$ 
(c.f. equation~\ref{bias-taylor}).  
Note that this expression assumes that the 2-halo--2-patch term is 
given simply by weighting the patch-patch correlation by the halo 
abundance within a patch.  

To a first approximation, patches do not overlap with one another, 
so the 2-halo--2-patch term should drop on scales smaller than the 
diameter of a patch.  It is on these scales that $P_{2h-1p}$ should 
begin to dominate.  
A first approximation for the net effect of $P_{2h-1p}$, then, is to 
not enforce this small-scale decrease of $P_{2h-2p}$, and to simply 
use the expression above for $P_{2h-2p}$ but $P_{\rm Lin}(k)$ 
instead of $P_{\rm Lin}(k|R_p)$, for all $k$.  
Section~\ref{BMVintegral} shows that this expression reduces to the 
standard two-halo term when $M_{min}\rightarrow 0$ and 
$M_{max}\rightarrow\infty$, so it is at least a reasonable 
approximation.  However, we will see below that the correlation 
function of objects in underdense regions sometimes shows a clear 
signature of the fact that patches do not overlap.  Thus, a more 
sophisticated approximation is required to describe clustering in 
underdense regions.  

To estimate the 2-halo--1-patch term, it is convenient to think of 
the patches as haloes, and of the haloes as patch-substructure.  
\cite{shethjain03} have developed the halo model description of 
clustering when haloes have substructure.  They allow for the 
possibility that a halo may be made up of a smooth component plus a 
population of subclumps.  Our present case corresponds to assigning 
all the mass to subclumps, and none to the smooth component, so that 
only the final two of the four terms in their equation~(31) contribute.  
Our expression for $P_{1h}(k|\delta)$ is effectively the same as their 
final term, so their third term is our $P_{2h-1p}(k|\delta)$.  Namely, 
\begin{equation}
 \begin{split}
   P_{2h-1p}(k|\delta) &= \int_{M_{min}}^{M_{max}} dM\, n(M,V)\\ 
    \ & \ \times \int_0^M dm_1 \,
         \left(\frac{m_1}{\bar{\rho_\delta}}\right) |u(k|m_1)| \\
    \ & \ \times \int_0^{M-m} dm_2 \,
         \left(\frac{m_2}{\bar{\rho_\delta}}\right) |u(k|m_2)| \\
    \ & \ \times \   N(m_1,m_2|M,V) \ U(k|m_1,m_2,M)^2.
 \end{split}
 \label{Pk2h1p}
\end{equation}
Here $U$ denotes the normalized Fourier transform of the spatial 
distribution of $m_1$ and $m_2$ haloes within a patch.  
A simple first estimate would use the correlation function of the 
haloes to model this profile, but to truncate this at the patch 
radius, since both haloes are required to lie within the patch:  
\begin{equation}
 U^2 \approx b(m_1)b(m_2)\,{
             P_{\rm Lin}(k) - P_{\rm Lin}(k|R_p)\over M/\bar\rho}
\end{equation}
where $P_{\rm Lin}(k)-P_{Lin}(k|R_p)$ denotes the power spectrum 
associated with setting the linear theory correlation function to 
zero on scales larger than the diameter of a patch $2R_p$.  
The other term, $N(m_1,m_2|M,V)$, denotes the average number of 
haloes of mass $m_1$ and $m_2$ in patches $V$ which contain total 
mass $M$.  \cite{sheth99} argue that, as a consequence of 
mass conservation, 
\begin{equation}
 N(m_1,m_2|M,V) \approx  N(m_1|M,V)\,N(m_2|M-m_1,V-v_{m_1}),
 \label{sl99}
\end{equation}
where $v_m$ is the volume associated with an $m$ halo.  

When $M\gg m$ and $V\gg v_m$, as it is in large overdense regions, 
then $N(m_1,m_2|M,V) \approx  N(m_1|M,V)\,N(m_2|M,V)$
i.e, $N(m_1,m_2|M,V)$ is well approximated by the product of the 
individual mean values.  In this case, 
\begin{equation}
 \begin{split}
   {P_{2h-1p}(k|\delta)\over P_{\rm Lin}(k) - P_{\rm Lin}(k|R_p)} \approx & 
    \int_{M_{min}}^{M_{max}} dM\, {n(M,V)\over M/\bar\rho}\\ 
    \ & \times \Biggl[\int_0^M dm \,N(m|M,V)\,b(m)\, \\
    \ & \qquad\quad 
        \times\left(\frac{m}{\bar{\rho_\delta}}\right) |u(k|m)|\Biggr]^2.
 \end{split}
 \label{slapprox}
\end{equation}
In underdense regions, however, simply using the product of the 
individual mean values is expected to be a bad approximation.  

The smooth curves in Figure~\ref{gifdm} show that the model developed 
above provides a good description of the environmental dependence of 
the dark matter correlation function.  In our comparisons, we have 
found that the full expression (equations~\ref{Pk2h1p}--\ref{sl99}) 
provides a substantially better description of $P_{2h-1p}$ in the 
underdense regions, whereas the simpler approximation of 
equation~(\ref{slapprox}) is adequate for the dense regions.  

\subsection{From dark matter to galaxies}
We now discuss how the model above can be extended to describe the 
environmental dependence of galaxy clustering.  
When all environments have been averaged over, the mean number density 
of galaxies $\bar n_{gal}$ is given by replacing the weighting by $m$ 
in equation~(\ref{rhobar}) for the mean density by 
$\langle N_{gal}|m\rangle$, the mean number of galaxies in an $m$-halo.  
For instance, one could use equation~(\ref{Ngsdss}) to write 
$\langle N_{gal}|m\rangle = 1 + \langle N_s|m\rangle$.  
Similarly, the weighting by $m/\bar\rho$ in $P_{2h}(k)$ is replaced with 
a weighting by $\langle N_{gal}|m\rangle/\bar n_{gal}$. 
And the weighting by $(m/\bar\rho)^2$ in the 1-halo term becomes 
$[2\langle N_s|m\rangle\,u(k|m) 
 + \langle N_s|m\rangle^2\, |u(k|m)|^2]/n_{gal}^2$.   
This weighting assumes there is always one galaxy at the centre, and 
that the number of satellite galaxies in an $m$-halo follows a Poisson 
distribution with mean $\langle N_s|m\rangle$.  

If one had an estimate of the dark matter density field, then one 
could include the environmental dependence of galaxy clustering by 
writing the mean number of galaxies in an $m$-halo surrounded by a 
region $V$ containing mass $M$ as $\langle N_{gal}|m,M,V\rangle$.  
If this mean number did not depend on $M$ and $V$, then the environmental 
dependence of galaxy clustering would be described by making the same 
replacements in the expressions for $P(k|\delta)$ as one makes for $P(k)$.  

In practice, one observes galaxies, not dark matter, so one has an 
estimate of the galaxy density field $\delta_{gal} = N/\bar n_{gal}V - 1$, 
and not of the dark matter $\delta = M/\bar\rho V -1$.  
Our previous expressions show what to do if the environment as 
defined by the mass density $\delta$ is known; 
describing galaxy clustering as a function of environment defined 
by the galaxies themselves, $\delta_{gal}$ rather than 
by the dark matter $\delta$, is considerably more complicated.  

\subsubsection{Effects of scatter in the $\delta_{gal}-\delta$ relation}
However, a simple approximate model can be derived if one assumes that 
$\langle N_{gal}|m,M,V\rangle = \langle N_{gal}|m\rangle$ is a monotonic 
function of $m$, and that the scatter around this mean relation is 
small.  The reason why is particularly easy to see if 
$\langle N_{gal}|m\rangle\propto m$ and there is no scatter around 
this relation.  If the environment of each galaxy is quantified by the 
number $N$ of other galaxies in a volume $V$ around it, then 
rank ordering cells by $N$, which is an observable, is the same as 
rank ordering by $M$, which is not.  
This rank ordering allows us to describe the environmental dependence 
of galaxy clustering by making simple adjustments to the expressions 
of Section~\ref{modelI}.  

In particular, suppose that one has measured how the number density of 
galaxies surrounded by regions containing at least $N_{min}$ other 
galaxies depends on $N_{min}$.  This number density is given by summing 
over the observable quantity $N$.  However, this number density can also 
be written as 
\begin{equation}
 \begin{split}
 \bar n_{\delta-gal} &= \int_{M_{min}(N_{min})}^\infty \!\!\!\! dM\,n(M,V)\\
      \ & \qquad\times\quad  \int_0^M dm\,N(m|M,V)\,\langle N_{gal}|m\rangle , 
 \end{split}
\end{equation}
where $M_{min}(N_{min})$ is obtained by requiring that the value of 
this expression match that observed as $N_{min}$ is varied.  
Once $M_{min}(N_{min})$ is known, the two-halo two-patch term can be 
written as 
\begin{eqnarray}
 \begin{split}
  P_{2h-2p}^{gal}(k|\delta) &\approx b_{\delta-gal}^2\,P_{Lin}(k),
                        \quad{\rm where}\\
  \bar n_{\delta-gal}\,b_{\delta-gal} &= 
      \int_{M_{min}}^\infty dM\,n(M,V)\,B(M,V) \\ 
  \ & \ \times\ \int_0^M dm\,N(m|M,V)\,\langle N_{gal}|m\rangle ,
 \end{split}
\end{eqnarray}
with the analagous substitutions for the $(m/\bar\rho_\delta)$ terms 
in equation~(\ref{Pk2h1p}) for $P_{2h-1p}(k|\delta)$.  
For similar reasons, the one-halo term in the centre plus Poisson 
satellites model is 
\begin{equation}
 \begin{split}
   P_{1h}^{gal}(k|\delta) = & \int_{M_{min}}^\infty dM\, n(M,V)
        \int_0^M dm \,N(m|M,V)\,\\ 
        \, & \times\ 
          {[2\langle N_s|m\rangle\,u(k|m)+\langle N_s|m\rangle^2\,|u(k|m)|^2]
           \over\bar n_{\delta-gal}^2} .
 \end{split}
\end{equation}
These expressions, which follow from those in Section~\ref{modelI}, 
are only accurate if the relation between the number of galaxies in a 
cell $N$ (the observable) and the mass in the cell $M$ is deterministic 
(i.e., has no scatter) and monotonic.  There {\em will} be scatter in 
$N$ at fixed $M$ if $\langle N_{gal}|m\rangle$ is a monotonic but 
nonlinear function of $m$, even if there is no scatter around the 
$\langle N_{gal}|m\rangle$ relation.  This scatter arises from the 
fact that there is scatter in the halo distribution at fixed $M$.  
(Write $N = \sum_i N(m_i)$ and $\sum_i m_i\equiv M$.  This shows 
that $N$ is independent of the distribution of the $m_i$ only if 
$N(m_i)\propto m_i$.)  

\begin{figure}
 \centering
 \includegraphics[width=\hsize]{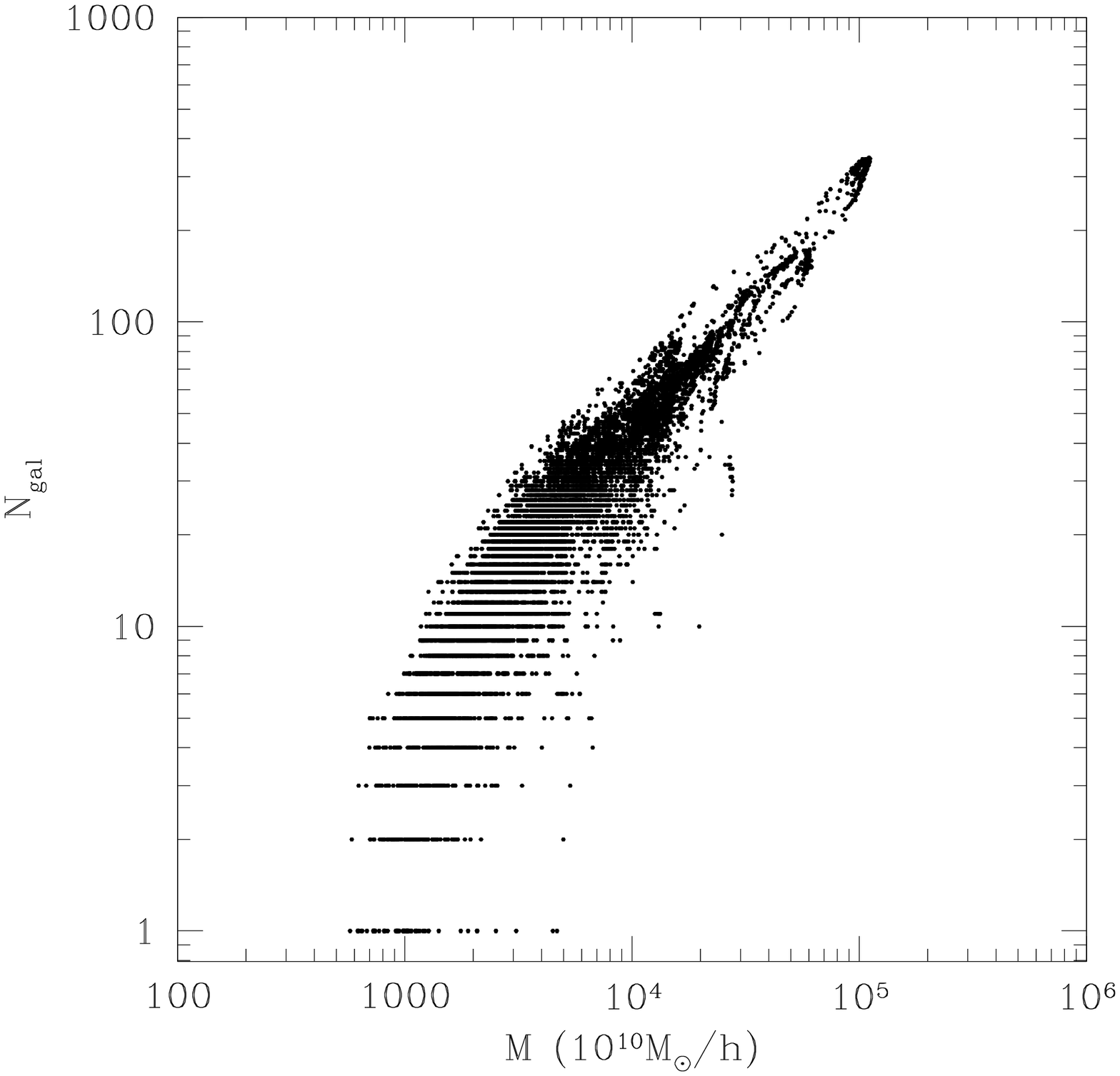}
 \caption{Number of galaxies within spheres of radii $5h^{-1}$~Mpc 
       centered on each galaxy, versus the total mass in the sphere, 
       for the galaxy model described in Section~\ref{mockgals}.  
       The small scatter in $M$ at large $N$ suggests that our model 
       should be reasonably accurate.}
 \label{NgMpatch}
\end{figure}

Scatter in $N$ at fixed $M$ means there is scatter in $M$ at fixed $N$.  
The expressions above should provide reasonable approximations to the 
exact description if the mean mass $M$ associated with a given value 
of $N$ in a cell, $\langle M|N\rangle$, is a monotonic function with 
small scatter.  In this context, `small' means that the dependence of 
clustering on environment (e.g., the mix of haloes) does not change 
dramatically over the range of environments associated with the rms 
scatter in $M$ around the mean $\langle M|N\rangle$.  
To see what this means, recall that, if the scale on which the
environment is defined is large, then equation~(\ref{nmdelta}) indicates 
that this dependence is proportional to $b(m)(M/\bar\rho V - 1)$.   
Massive haloes have $b(m)>1$, with $b(m)$ a strongly increasing 
function of $m$, whereas low mass haloes have $b(m)<1$, and the 
$m$-dependence is weak.  Since the most massive haloes populate 
the densest cells, the model can tolerate larger scatter in the 
$M-N$ relation at small $N$ than at large $N$.  

Figure~\ref{NgMpatch} shows the relation between the number of 
galaxies $N$ and the mass $M$ within a sphere of radius $5h^{-1}$Mpc 
centred on each galaxy, for the galaxy model described in 
Section~\ref{mockgals}.  The figure indicates that treating $N$ as a 
monotonic deterministic function of $M$ (and vice-versa) is a good 
approximation at least at high masses, even though the underlying 
relation between number of galaxies and halo mass, 
$\langle N_{gal}|m\rangle$, is nonlinear (equation~\ref{Ngsdss}).  
Notice that the scatter around the mean $\langle M|N\rangle$ relation 
is particularly small at large $N$.  Although the scatter increases 
at smaller $N$, we argued above that, in this regime, the effect of 
scatter is less important.  
Hence, Figure~\ref{NgMpatch} suggests that the simple model developed 
in this section should be reasonably accurate.  

Recall that Figure~\ref{gifsdss} shows measurements of the 
environmental dependence of galaxy clustering made in a mock galaxy 
catalog constructed using the GIF simulation following methods 
outlined in Section~\ref{mockgals}.  
The smooth curves in Figure~\ref{gifsdss} show that the model 
developed here is in reasonable agreement with the measured 
dependence of $\xi(r)$ on $\delta_{gal}$.

\subsubsection{The two contributions to the 2-halo term}
In constructing the model, we remarked that there were two types of 
contribution to the two-halo term.  So one might wonder if there is 
some clear signature of the transition from one type of contribution 
to another.  The mock galaxy sample used for Figure~\ref{gifsdss} 
does not result in a $\xi(r|\delta)$ with a clear inflection or break 
on the patch scale.  

\begin{figure}
\centering
 \includegraphics[width=\hsize]{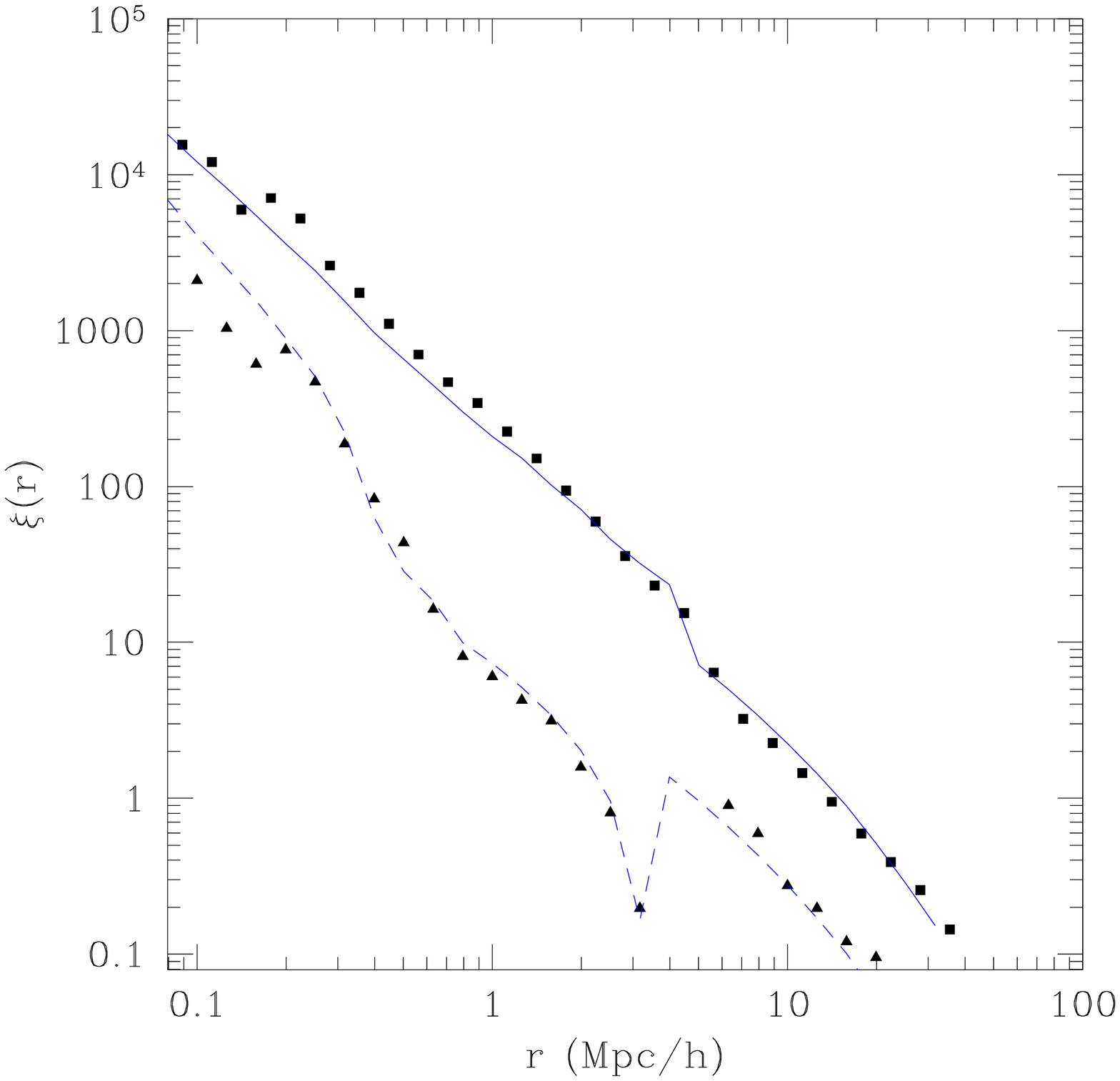}
 \includegraphics[width=\hsize]{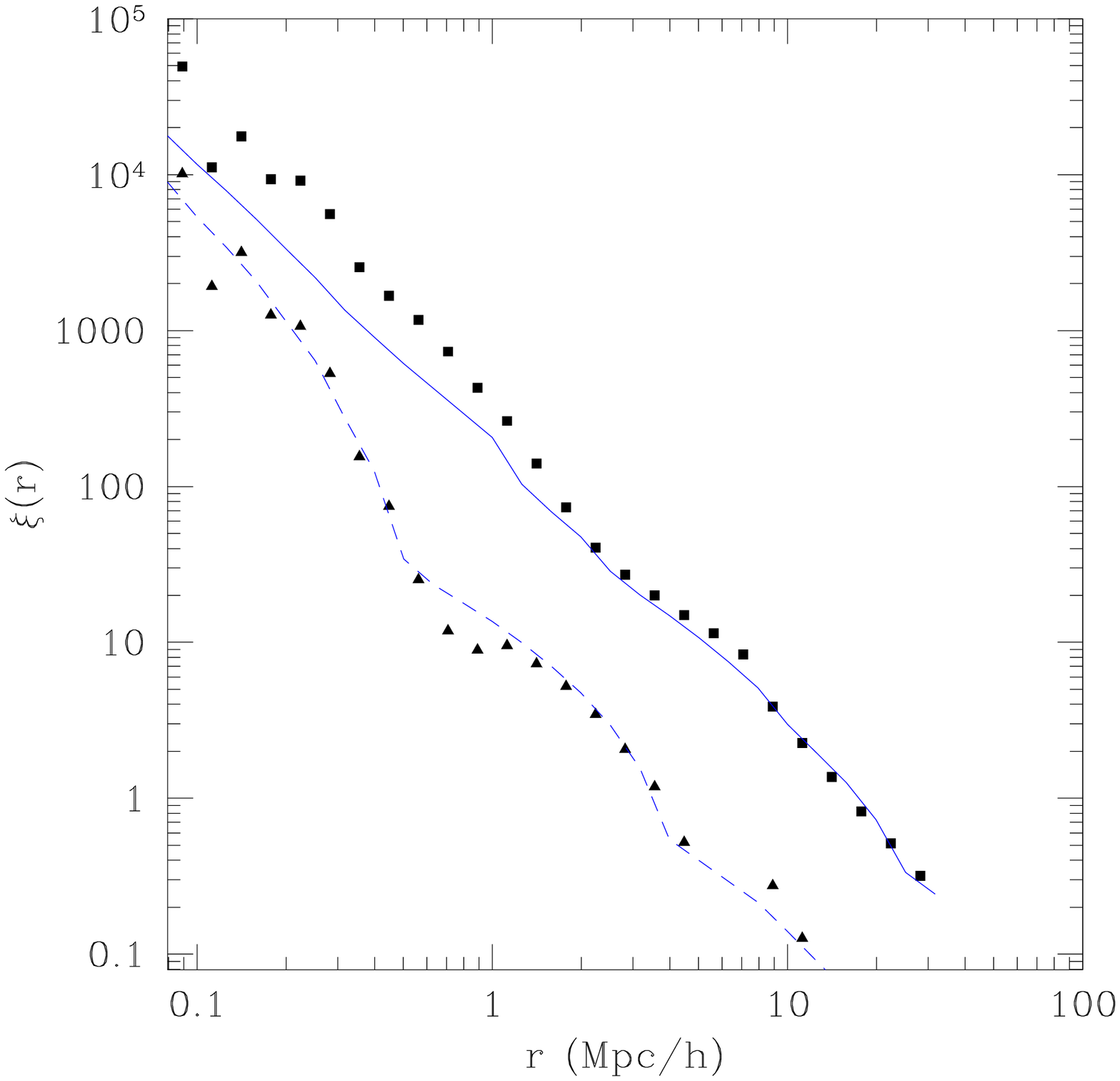}
 \caption{Correlation function for mock galaxies in the VLS simulations. 
  Squares show $\xi(r|\delta)$ for the overdense sample 
  and triangles are for the underdense sample.
  In the top panel, the density was defined by counting galaxies 
  within spheres having radius $5h^{-1}$Mpc centred on each object, 
  and the bottom panel used $8h^{-1}$Mpc spheres.  
  Curves show the associated analytic model which is able to describe 
  the three different clustering regimes. 
  }
 \label{vlssdss}
\end{figure}

However, such a transition is seen in Figure~\ref{vlssdss}, which 
shows results for a mock galaxy sample in the VLS simulations, 
constructed with $M_{min}= 10^{12.72} h^{-1}M_\odot$ and $\alpha=1.39$, 
chosen to be similar to galaxies more luminous than $M_r=-21$.  
The top and bottom panels show results from analyses 
in which the large scale environment was defined by the mass within 
spheres of radii $5h^{-1}$ and $8h^{-1}$Mpc, respectively.  
The inflection indicating the transition from $\xi_{2h-1p}(r|\delta)$ 
to $\xi_{2h-2p}(r|\delta)$ is more clearly seen for low density regions, 
and it is clearer when the scale which defines the environment is 
smaller.  For these more luminous galaxies, using the simpler 
equation~(\ref{slapprox}) in $P_{2h-1p}$ is an excellent approximation 
in dense regions, but grossly overestimates (by at least a factor of two) 
the clustering strength on intermediate scales in underdense regions.  
In underdense regions, equations~(\ref{Pk2h1p})--(\ref{sl99}) are 
substantially more accurate.  
We would like to point out that the relation between $N$ and $M$ 
(that was mentioned in the previos section) within spheres of $8h^{-1}$Mpc,
shows more scatter at large $N$. As the model is more sensitive to larger
scatter at large $N$, we believe this causes the analytical prediction
to depart from the simulation results at small scales, 
as shown in the lower panel of Figure~\ref{vlssdss}. 

Figure~\ref{gifdm} indicates that our model for the environmental 
dependence of dark matter clustering is accurate, and 
the agreement between the symbols and the curves in 
Figures~\ref{gifsdss} and~\ref{vlssdss} indicates that our approximate 
model for the environmental dependence of galaxy clustering also works 
quite well.  Although it is reassuring that the model works so well---it 
suggests that we understand the physics which drives the environmental 
dependence of clustering---this analytic description is not needed to 
perform the test of environmental effects described in the main text.

\subsection{Consistency checks}\label{BMVintegral}
This section shows that the expressions in Section~\ref{modelI} 
do reduce to the standard expressions in Section~\ref{standard} 
upon averaging over all environments.  

The mass density is 
\begin{equation}
 \begin{split}
  \bar{\rho}_\delta &= \int_{M_{min}}^{M_{max}} dM\,n(M,V) 
                          \int_0^M dm\, N(m|M,V)\,m \\
     &\to \int_0^\infty dm\,m \int_m^\infty dM\,n(M,V)\,N(m|M,V) \\
     &= \int_0^\infty dm\,{dn(m)\over dm}\,m\int_m^\infty dM\,f(M,V|m)\\
     &= \int_0^\infty dm\,{dn(m)\over dm}\,m =  \bar\rho
 \end{split}
\end{equation}
when $M_{min}\to 0$ and $M_{max}\to\infty$ (this limit corresponds to 
averaging over all environments).  
In this limit, the one-halo term is
\begin{eqnarray}
   P_{1h}(k|\delta) &\to& \int_0^\infty dM\, n(M,V)
          \nonumber\\
   &&\quad\times\quad \int_0^M dm \, N(m|M,V)
       \left({m\over \bar\rho}\right)^2 |u(k|m)|^2 \nonumber\\
   &=& \int_0^\infty dm\,\left({m\over \bar\rho}\right)^2\, |u(k|m)|^2 
       \nonumber\\
   &&\qquad\times\quad \int_m^\infty dM\,n(M,V)\, N(m|M,V)
     \nonumber\\
   &=& \int_0^\infty dm\, {dn(m)\over dm}\,
         \left({m\over \bar\rho}\right)^2\, |u(k|m)|^2 .
\end{eqnarray}
Similarly, the two halo term becomes 
\begin{eqnarray}
 {P_{2h-2p}(k|\delta)\over P_{Lin}(k|R_p)} &\to& 
    \Biggl[\int_0^\infty dM\, n(M,V)\,B(M,V)\nonumber\\
   && \times \int_0^M dm \,N(m|M,V)
        \left({m\over\bar\rho}\right)\,u(k|m)\Biggr]^2 \nonumber\\
   &=& \Biggl[\int_0^\infty dm\,{m\over\bar\rho}\, u(k|m)
       \int_m^\infty dM\,N(m|M,V)\nonumber\\
   && \qquad\qquad \times \quad n(M,V)\,B(M,V)\Biggr]^2.
  \label{P2h2p}	
\end{eqnarray}
This can be simplified as follows.  The halo overdensity is 
\begin{equation}
  \delta_h(m|M,V) = {dN(m|M,V)/dm\over V\,dn(m)/dm} - 1 
                 \equiv b(m|M,V)\, \delta.
\end{equation}
When $V\gg 1$, then $\delta\ll 1$, and $M$ is almost surely much larger 
than the typical halo mass, so $M\gg m$ for most values of $m$.  
In this limit, 
\begin{eqnarray}
 \begin{split}
 {dN(m,\delta_c|M,V)/dm\over V\,dn(m)/dm} &= 
  (1+\delta)\, {dn[m,\delta_c-\delta_0(\delta)]/dm\over dn(m)/dm}\\
  \ &\to (1+\delta)\,
    \left(1 - \delta_0(\delta){d\ln dn(m)/dm\over d\delta_c}\right)\\
  \ &\to 1+\delta - \delta\,{d\ln dn(m)/dm\over d\delta_c},
 \end{split}
\end{eqnarray} 
where we have used the fact that $\delta_0\approx\delta\ll 1$.  
Hence, we can approximate 
$\delta_h(m)\approx b(m)\delta$ where 
 $b(m)\approx 1 - d\ln n(m,\delta_c)/d\delta_c$
(equation~\ref{bias-taylor}) is no longer a function of $V$.  
Our assumption is that $B(M)$ is related to $n(M,V)$ similarly.  
Thus, the second integral in the expression above becomes 
\begin{eqnarray}
 \! && \int_m^\infty dM\, n(M,V)\, B(M,V)\, N(m|M,V)\nonumber\\
 &=& \int_m^\infty dM\, n(M,V)\, 
       \left[1-{d\ln n(M,V)\over d\delta_0}\right]\, N(m|M,V)\nonumber\\
 &=& {d n(m)\over dm} - \int_m^\infty dM\,
       {d n(M,V)\over d\delta_0}\, N(m|M,V)\nonumber\\
 &=& {d n(m)\over dm} - {d\over d\delta_0}\int_m^\infty dM\,
       n(M,V)\, N(m|M,V)\nonumber\\ 
 &&\qquad + \int_m^\infty dM\,n(M,V)\, {dN(m|M,V)\over d\delta_0}\nonumber\\
 &=& {d n(m)\over dm} - {d n(m)\over d\delta_0} 
       - {d\over d\delta_c}\int_m^\infty dM\, n(M,V)\, N(m|M,V)\nonumber\\ 
  &=& {d n(m)\over dm}\left[1 - {d \ln dn(m)/dm\over d\delta_c}\right]\nonumber\\
 &=& {d n(m)\over dm}\,b(m).
\end{eqnarray}
Hence 
\begin{equation}
 {P_{2h-2p}(k|\delta)\over P_{Lin}(k|R_p)} 
   \to \Biggl[\int_0^\infty dm\, {dn(m)\over dm}\,{m\over\bar\rho}\,
        b(m)\,u(k|m)\Biggr]^2.
\end{equation}
If we do not truncate the two-halo term on small scales, as a crude 
approximation for $P_{2h-1p}$, i.e., if we simply set 
 $P_{Lin}(k|R_p)\to P_{Lin}(k)$, 
then this agrees with equation~(\ref{Pk1h2h}).  

If we do truncate, then we must check if the large-scale limit 
of $P_{2h-1p} + P_{2h-2p}$ agrees with equation~(\ref{Pk1h2h}).
In the limit of large patches and all environments, the two-halo 
one-patch term becomes 
\begin{eqnarray*}
  && {P_{2h-1p}(k|\delta)\over P_{\rm Lin}(k) - P_{\rm Lin}(k|R_p)} \to 
    \int_0^\infty dM\, {n(M,V) (1+\delta)\over M/\bar\rho}\\
    && \times \left[\int_0^M dm \,{dn(m)\over dm}\,
              {m\over\bar\rho}\,b(m)\,u(k|m)\,V[1+b(m)\delta]\right]^2.
\end{eqnarray*}
On large scales $\delta\ll 1$ so we can ignore $b\delta$ compared 
to unity.  Since $M/\bar\rho = V(1+\delta)$, 
\begin{eqnarray}
  && {P_{2h-1p}(k|\delta)\over P_{\rm Lin}(k) - P_{\rm Lin}(k|R_p)} \approx 
    \int_0^\infty dM\, Vn(M,V)\nonumber\\
    && \qquad\qquad\qquad\qquad \times 
         \left[\int_0^M dm \,{dn(m)\over dm}\,\frac{m}{\bar\rho}\,
         b(m)\,u(k|m)\right]^2\nonumber\\
  && \qquad\qquad \approx \int_0^\infty dM\, Vn(M,V)\nonumber\\
    && \qquad\qquad\qquad\qquad \times 
         \left[\int_0^\infty dm \,{dn(m)\over dm}\,\frac{m}{\bar\rho}\,
         b(m)\,u(k|m)\right]^2\nonumber\\
  && \qquad\qquad = 
     \left[\int_0^\infty dm \,{dn(m)\over dm}\,\frac{m}{\bar\rho}\,
         b(m)\,u(k|m)\right]^2
\end{eqnarray}
where the second line follows from the fact that if $M$ is much 
larger than the mass of the largest halo, then the upper limit 
of the integrals over $m$ can safely be changed from $M$ to 
$\infty$, and the final line uses the fact that the integral over 
$M$ which remains is the same as the integral over the 
counts-in-cells probability distribution, and so equals unity.  
This shows explicitly that, in this limit, $P_{2h-1p} + P_{2h-2p}$ 
agrees with equation~(\ref{Pk1h2h}).

\label{lastpage}


\begin{thebibliography}{99}
\bibitem[\protect\citeauthoryear{Balogh \& Morris}{2000}]{balogh00} 
 Balogh M.L., Morris S.L., 2000, MNRAS, 318, 703
\bibitem[\protect\citeauthoryear{Balogh et al.}{2002}]{balogh02} 
 Balogh M.L. et al., 2002, MNRAS, 337, 256
\bibitem[\protect\citeauthoryear{Balogh et al.}{2004}]{balogh04}
 Balogh M.L. et al., 2004, MNRAS, 348, 1355 
\bibitem[\protect\citeauthoryear{Benson et al.}{2001}]{benson01}
 Benson A.J. et al., 2001, MNRAS, 327, 1041 
\bibitem[\protect\citeauthoryear{Blanton et al.}{2004}]{blanton04}
 Blanton M.R., Eisenstein D.J., Hogg D.W., Zehavi I., 2004, astro-ph/0411037
\bibitem[\protect\citeauthoryear{Cole et al.}{2000}]{cole00} 
 Cole S., Lacey C.G., Baugh C.M., Frenk C.S., 2000, MNRAS, 319, 168
\bibitem[\protect\citeauthoryear{Cooray \& Sheth}{2002}]{cooray02} 
 Cooray A., Sheth R.K., 2002, Phys. Rep., 372, 1 
\bibitem[\protect\citeauthoryear{Dressler}{1980}]{dressler80} 
 Dressler A., 1980, ApJ, 236, 351
\bibitem[\protect\citeauthoryear{Farouki \& Shapiro}{1980}]{farouki80} 
 Farouki R., Shapiro S.L., 1980, ApJ, 241, 928
\bibitem[\protect\citeauthoryear{Gao, Springel \& White}{2005}]{gao05} 
 Gao L., Springel V., White S. D. M., 2005, MNRAS, submitted
\bibitem[\protect\citeauthoryear{Gomez et al.}{2003}]{gomez03} 
 Gomez P.L. et al., 2003, ApJ, 584, 210
\bibitem[\protect\citeauthoryear{Gunn \& Gott}{1972}]{gunn72} 
 Gunn J.E., Gott J.R.I., 1972, ApJ, 528, 118
\bibitem[\protect\citeauthoryear{Hogg et al}{2004}]{hogg04} 
 Hogg D.W. et al., 2004, ApJ, 601, L29 
\bibitem[\protect\citeauthoryear{Kaiser}{1984}]{kaiser84} 
 Kaiser N., 1984, ApJ, 284, 9
\bibitem[\protect\citeauthoryear{Kauffmann et al.}{1997}]{kauffmann97} 
 Kauffmann G., Nusser A., Steinmetz M., 1997, MNRAS, 286, 795
\bibitem[\protect\citeauthoryear{Kauffmann et al.}{1999}]{kauffmann99} 
 Kauffmann G., Colberg J.M., Diaferio A., White S.D.M., 1999, MNRAS, 307, 529 
\bibitem[\protect\citeauthoryear{Kauffmann et al.}{2004}]{kauffmann04} 
 Kauffmann G. et al., 2004, MNRAS, 353, 713 
\bibitem[\protect\citeauthoryear{Kravtsov et al.}{2004}]{kravtsov04} 
 Kravtsov A. et al., 2004, ApJ, 609, 35
\bibitem[\protect\citeauthoryear{Lacey \& Cole}{1993}]{lacey93} 
 Lacey C., Cole S., 1993, MNRAS, 262, 627
\bibitem[\protect\citeauthoryear{Lemson \& Kauffmann}{1999}]{lemson99} 
 Lemson G., Kauffmann G., 1999, MNRAS, 302, 111
\bibitem[\protect\citeauthoryear{Mo \& White}{1996}]{mowhite96} 
 Mo H.J., White S.D.M., 1996, MNRAS, 282, 347 
\bibitem[\protect\citeauthoryear{Mo et al.}{2004}]{mo04} 
 Mo H. J., Yang X., van den Bosch F. C., Jing Y. P., 2004, MNRAS, 349, 205
\bibitem[\protect\citeauthoryear{Moore et al.}{1996}]{moore96}
 Moore B., Katz N., Lake G., Dressler A., Oemler A., 1996, Nat, 379, 613
\bibitem[\protect\citeauthoryear{Navarro et al.}{1997}]{navarro97} 
 Navarro J.F., Frenk C.S., White S.D.M., 1997, ApJ, 490, 493
\bibitem[\protect\citeauthoryear{Peebles}{1974}]{peebles74} 
 Peebles P.J.E., 1974, ApJ, 189, L51
\bibitem[\protect\citeauthoryear{Peebles}{1980}]{peebles80} 
 Peebles P.J.E., 1980, The Large-Scale Structure of the Universe.
           Princeton Univ. Press, Princeton, NJ
\bibitem[\protect\citeauthoryear{Press \& Schechter}{1974}]{press74} 
 Press W.H., Schechter P., ApJ, 1974, 187, 425
\bibitem[\protect\citeauthoryear{Sheth}{1998}]{sheth98} 
 Sheth R.K., 1998, MNRAS, 300, 105
\bibitem[\protect\citeauthoryear{Sheth \& Jain}{1997}]{shethjain97} 
 Sheth R.K., Jain B., 1997, MNRAS, 285, 231
\bibitem[\protect\citeauthoryear{Sheth \& Jain}{2003}]{shethjain03} 
 Sheth R.K., Jain B., 2003, MNRAS, 345, 62
\bibitem[\protect\citeauthoryear{Sheth \& Lemson}{1999}]{sheth99} 
 Sheth R.K., Lemson G., 1999, MNRAS, 304, 767
\bibitem[\protect\citeauthoryear{Sheth \& Tormen}{1999}]{shethtor99} 
 Sheth R.K., Tormen G., 1999, MNRAS, 308, 119
\bibitem[\protect\citeauthoryear{Sheth \& Tormen}{2002}]{shethtor02} 
 Sheth R.K., Tormen G., 2002, MNRAS, 329, 61 
\bibitem[\protect\citeauthoryear{Sheth \& Tormen}{2004}]{shethtor04} 
 Sheth R.K., Tormen G., 2004, MNRAS, 350, 1385
\bibitem[\protect\citeauthoryear{Sheth, Abbas \& Skibba}{2004}]{shethabbas04} 
 Sheth R.K., Abbas U., Skibba R.A., in Diaferio A., ed, 2004, 
           Proc. IAU Coll. 195, 
           Outskirts of galaxy clusters: intense life in the suburbs, 
           CUP, Cambridge, p.~349
\bibitem[\protect\citeauthoryear{Sheth, Mo \& Tormen}{Sheth et al. 2001}{2001}]{shethmot01} 
 Sheth R.K., Mo H., Tormen G., 2001, MNRAS, 323, 1
\bibitem[\protect\citeauthoryear{Sheth et al.}{2001}]{sheth01} 
 Sheth R.K., Hui L., Diaferio A., Scoccimarro R., 2001, MNRAS, 325, 1288
\bibitem[\protect\citeauthoryear{Somerville \& Primack}{1999}]{somerville99} 
 Somerville R., Primack J.R.S., 1999, MNRAS, 310, 1087
\bibitem[\protect\citeauthoryear{White \& Frenk}{1991}]{white91}
 White S.D.M., Frenk C.S., 1991, ApJ, 379, 52
\bibitem[\protect\citeauthoryear{White \& Rees}{1978}]{white78}
 White S.D.M., Rees M.J., 1978, MNRAS, 183, 341
\bibitem[\protect\citeauthoryear{Yang et al.}{2003}]{yang03} 
 Yang X., Mo H.J., van den Bosch F.C., 2003, MNRAS, 339, 1057
\bibitem[\protect\citeauthoryear{Zehavi et al.}{2004}]{zehavi04} 
 Zehavi I. et al., 2004, astro-ph/0408569
\end{thebibliography}
\end{document}